\documentstyle[fleqn,epsf]{tp}
\begin{document}

\twocolumn[
\title{Broken Defects}
\author{Ue-Li Pen$^1$\\
{\it $^1$CITA, 60 St. George St., Toronto, M5S 3H8, Canada}}
\vspace*{16pt}   

ABSTRACT.\
Recent indications of a neutrino mass raise the possibility that the
dark matter may be hot, which would be a severe challenge for
structure formation theory.  We argue that generic defect theories fare
very well in hot dark matter cosmologies, and also work in the
presence of curvature or a cosmological constant.
If this model is correct, then the MAP
and PLANCK missions will not measure what people expect them to
(oscillations); rather, they will measure a broad hump.
\endabstract]

The traditional class of scale invariant defect models has severe
problems\cite{watson97,pst97}.  They do not explain the rise in
fluctuation amplitude at scales of a few degrees, and the slope of the
perturbations is at conflict with the observed matter power spectrum.
We show here that global topological defects generically change
behaviour at the matter-radiation transition at a cosmic age of 100000
years.  In this case, instead of being at great discrepancy with the
cosmic microwave background radiation fluctuations, this model in fact
correctly explains several hitherto mysterious phenomena.  Recent
measurements have indicated that neutrinos have mass, and have opened
up that the dark matter may be massive neutrinos.  If we accept that
neutrinos have mass, we have the potential consequence that neutrinos
account for all the dark mass in the universe.  This simple scenario
is consistent with the simplest massive neutrino extensions of the
standard particle physics models\cite{georgi}, and challenges
cosmologists to contemplate its implications for structure formation.
The scale invariant adiabatic model predicted by inflation is not
consistent if massive neutrinos account for the bulk of gravitating
matter.  We will show that generic defect models would now become the
best physically based model of structure formation.  The most recent
supernovae search\cite{science,perlmutter}, concordance
arguments\cite{oststei} and direct angular diameter distance
measurements\cite{danos} have suggested that the universe is filled
with a cosmological constant.  This further challenges hot dark matter
(HDM) structure formation models, but will turn out to be simple to
reconcile in defect models.

Let us now address the dynamics of global defects at matter-radiation
equality.  In modern particle physics, all interactions which are not
forbidden actually occur.  On this ground, one generically expects
couplings between gravity and other quantum fields which are stronger
than just minimal coupling to the space time curvature.  Global fields
experience couplings of the form $\epsilon\phi_1^\dagger\phi_2 R$,
where $\epsilon$ is a dimensionless coupling constant of order unity,
and $R$ is the scalar curvature.  Einstein equations relate
$R=T=\rho-3p$.  In the radiation epoch of the universe, $R= 0$, so the
coupling has no dynamical effect.  But when the universe enters matter
domination, this coupling will generally modify the symmetry which
gave rise to the defects, and they can change their dynamics.  During
the onset of change, the vacuum energy contributed by the new coupling
will boost the power on the largest scales, making the power spectrum
consistent with the observed excess of large scale power.

These defect models have two free parameters.  One is the symmetry
breaking scale $\phi_0$, and the second is its coupling to the Riemann
scalar $\epsilon$.  From the velocity dispersion in clusters of
galaxies we expect $\phi_0/m_{\rm planck} \sim v^2/c^2 \sim 10^{-5}$,
which would also be related to the COBE fluctuations.  This energy
scale suggests that the defect would be relicts of the breaking of the
Grand-Unified-Theory in the early universe.  A predictive property of
this model is the non-Gaussianity of fluctuations.  The strongest
constraints arises from the statistics of rich clusters\cite{chiu97}.
Several clusters with temperatures above 12 keV have been
found\cite{donahue}, and a recent cluster with a temperature of 17
keV\cite{tucker97} has been discovered.  The existence of these very
hot clusters is becoming an enigma for any Gaussian cosmological
model.  In non-Gaussian defect models, one would expect a larger
proportion of high temperature clusters relative to low temperature
ones.  The recent Las Campanas galaxy survey discovered features in
the two dimensional galaxy power spectrum which is inconsistent with
Gaussianity\cite{landy}.  Such signs of non-Gaussianity are
generically predicted in these defect models.

Several qualititive phenomena may happen due to the non-minimal
coupling.  Because the field interactions are different in the
radiation and matter epochs, the defect energy density could be larger
in the radiation epoch, which would solve the large scale structure
problems of scaling defects.  The hot dark matter experiences some
damping out of small scale power, thus relieving the problem of excess
fluctuations on small scales.  Unlike the adiabatic case, the
perturbations on small scales do not all free stream and damp
expontially, but rather the power spectrum of the source perturbations
at matter-radiation equality will be imprinted on the matter power
spectrum.  In simulations we have observed that the onset of the phase
transition at matter radiation equality furthermore injects power at
the horizon scale at that time, or about 100 Mpc scales\cite{pt99}.

In the defect scenario, the small scale CMB experiments MAP and Planck
will not measure the multiple oscillations expected from adiabatic
perturbations imprinted before the big bang, as predicted by
inflation.  Instead, the degree scale anisotropies are caused by the
incoherent properties of global defects\cite{pst97}.  Without the
oscillations, it will be very difficult to reconstruct cosmological
parameters accurately.  On the other hand, this would be an exciting
measurement of fundamental physics and the history of symmetry
breaking.

\section*{Acknowledgements}.  The idea of broken scale invariance from
non-minimal coupling is due to Neil Turok.  I thank Uros Seljak for
helpful discussions.

\end{document}